\documentclass[pra,aps,10pt,twocolumn]{revtex4-1}

\usepackage{amsmath,amssymb}
\usepackage[utf8]{inputenc}
\usepackage{graphicx}
\graphicspath{{plots/}}

\newcommand{\tr}{\mathop{\mathrm{tr}}}

\begin{document}

\title{Dispersive readout: Universal theory beyond the rotating-wave approximation}
\author{Sigmund Kohler}
\affiliation{Instituto de Ciencia de Materiales de Madrid, CSIC, E-28049 Madrid, Spain}

\begin{abstract}
We present a unified picture of dispersive readout of quantum systems in
and out of equilibrium.  A cornerstone of the approach is the backaction of
the measured system to the cavity obtained with non-equilibrium
linear-response theory. It provides the dispersive shift of the cavity
frequency in terms of a system susceptibility.  It turns out that already
effortless computations of the susceptibility allow one to generalize
former results beyond a rotating-wave approximation (RWA). Examples are the
readout of detuned qubits and thermally excited multi-level systems.  For
ac-driven quantum systems, we identify the relevant Fourier component of
the susceptibility and introduce a computational scheme based on Floquet
theory.  The usefulness is demonstrated for two-tone spectroscopy and
interference effects in driven two-level systems.  This also reveals that
dispersive readout does not necessarily measure excitation probabilities.
\end{abstract}

\date{\today}
\maketitle

\section{Introduction}
\label{sec:intro}

An essential task in quantum information processing is the readout of the
final state of a the system.  For solid state qubits, this may be
performed by energy selective escape from a metastable potential
\cite{Martinis2002a, Hanson2005a} or with a bifurcation amplifier
\cite{Siddiqi2004a, Mallet2009a}.  A further established technique for this
aim is dispersive readout \cite{Blais2004a} which is based on the coupling
of the qubit to a superconducting transmission line, henceforth ``cavity''.
Owing to the interaction with the qubit, the cavity experiences a
frequency shift that depends on the qubit state.  This shift can be probed
experimentally via the cavity transmission and reflection.  The relation
between this response and the qubit state can be obtained by transforming
the qubit-cavity Hamiltonian to the dispersive frame \cite{Blais2004a}. The
calculation is usually performed within a rotating-wave approximation (RWA)
valid when the detuning of qubit and cavity is rather small, but still
larger than their mutual coupling.

Experimental progress motivated several generalizations such as a treatment
beyond RWA \cite{Zueco2009b}.  Recently, dispersive readout has been
proposed for multi-level systems within RWA \cite{Burkard2016a,
Benito2017a} and for ac-driven quantum systems \cite{Kohler2017a}.  A main
goal of the present work is to put these approaches to a common ground by
computing the backaction of the system to the cavity within non-equilibrium
linear response theory.  This will demonstrate that generically, the
dispersive shift is given by the auto-correlation function or
susceptibility of the system operator by which the coupling to the cavity
is established.  The fact that this susceptibility depends only on the
system and not on the cavity makes the approach universal and applicable to
a wide class of setups.  Moreover, it provides non-RWA corrections in a
straightforward and technically simple manner.

This work is organized as follows.  In Sec.~\ref{sec:model}, we introduce
the system-cavity model and derive with the input-output formalism
\cite{Collett1984a, Gardiner2004a, Blais2004a, Clerk2010a} the relation
between the cavity transmission and the response function of the system.
In Sec.~\ref{sec:undriven}, the theory is applied to single qubits and to
multi-level systems with a focus on non-RWA corrections.
Section~\ref{sec:ac} is devoted to the peculiarities of ac-driven systems,
while the conclusions are drawn in Sec.~\ref{sec:summary}.

\begin{figure}[b]
\centerline{\includegraphics{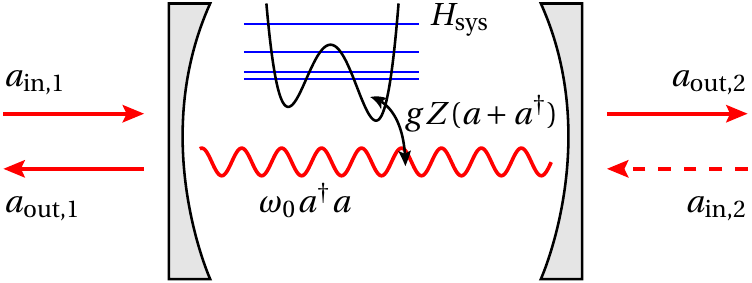}}
\caption{Cavity coupled to input and output modes, as well as to the
quantum system to be measured.  In the absence of the system, a symmetric
cavity ($\kappa_1=\kappa_2$) perfectly transmits resonant input at
$\omega=\omega_0$.  The input field on the right-hand side,
$a_{\text{in},2}$, is in its vacuum state and, hence, does not contribute to the
average output fields, but nevertheless causes dissipation.}
\label{fig:setup}
\end{figure}

\section{System-cavity model and cavity response}
\label{sec:model}

We consider the setup sketched in Fig.~\ref{fig:setup} with the quantum
system to be measured, e.g.\ a qubit, described by a still unspecified
and generally time-dependent Hamiltonian $H_\text{sys}$.  It interacts with
an open cavity such that the system-cavity Hamiltonian reads (in units with
$\hbar=1$)
\begin{equation}
H = H_\text{sys} + gZ(a^\dagger+a) + \omega_0 a^\dagger a
\label{Hsc}
\end{equation}
with the cavity frequency $\omega_0$ and the corresponding bosonic
operators $a$ and $a^\dagger$.
Owing to the coupling, the cavity acts upon the quantum system and in turn
experiences a backaction that shifts the cavity frequency.  This dispersive
shift or cavity pull is visible in the transmission and allows one to probe
the system.  The resolution is mainly determined by the cavity decay rate
$\kappa$.

The paradigmatic case is a qubit with $H_\text{sys} = \frac{\Delta}{2}
\sigma_z$ and dipole coupling $Z=\sigma_x$, written in the tunnel basis of
delocalized states.  If the cavity and the qubit are close to resonance
while the coupling constant $g$ is sufficiently small, the effective cavity
frequency changes as $\omega_0 \to \omega_0 +\delta\omega$ with the
dispersive shift $\delta\omega = \pm g^2/(\Delta-\omega_0)$
\cite{Blais2004a}; see Sec.~\ref{sec:undriven} where this result is
re-derived with the present formalism.  The sign $\pm$ corresponds to the
qubit states $|{\uparrow}\rangle$ and $|{\downarrow}\rangle$, respectively.
Quantitatively, the operating regime is
\begin{equation}
\kappa \ll g \ll |\omega_0-\Delta|
\label{readoutcondition}
\end{equation}
together with the RWA condition $|\omega_0-\Delta| \ll \omega_0
\approx\Delta$.  The second inequality in Eq.~\eqref{readoutcondition} is
used for the underlying perturbation theory \cite{Blais2004a, Zueco2009b}.
Together with the requirement that the width of the cavity resonance must
be smaller than the dispersive shift, $\kappa\lesssim
g^2/|\Delta-\omega_0|$, follows the first inequality.  This result for
$\delta\omega$ will emerge as the RWA limit of a special case.
Moreover, we will see that Eq.~\eqref{readoutcondition} can be replaced by
the weaker condition that the impact of the cavity on the system must be
within the linear response limit.

\subsection{Input-output theory}

A suitable tool for computing a cavity response is input-output theory
\cite{Collett1984a, Gardiner2004a} which
for the cavity provides the quantum Langevin equation \cite{Clerk2010a,
Petersson2012a, Burkard2016a}
\begin{equation}
\label{dota}
\dot a = -i\omega_0 a -igZ_t -\frac{\kappa}{2} a
-\sum_{\nu=1,2} \sqrt{\kappa_\nu} a_{\text{in},\nu} \,.
\end{equation}
Its first two terms are due to the Heisenberg equation of motion $-i[H,a]$,
while the dissipative term with the cavity loss rate $\kappa =
\kappa_1+\kappa_2 \equiv \omega_0/Q$ and the input field
$a_{\text{in},\nu}$ stem from the interaction with the electric circuit.
Possible further losses will augment $\kappa$, but are not considered
here.  The input field $a_{\text{in},1}$ may be monochromatic or broadband,
while $a_{\text{in},2}$ is in its vacuum state.  From the corresponding
time-reversed equation one finds the input-output relation
$a_{\text{out},\nu} -a_{\text{in},\nu} = \sqrt{\kappa_\nu}a$.  Since we are
not interested in quantum fluctuations of the cavity field, we consider
Eq.~\eqref{dota} in its classical limit as an equation of motion for the
expectation values $a_t \equiv \langle a\rangle_t$ and $Z_t \equiv \langle
Z\rangle_t$.

Our strategy is to express $Z_t$ in terms of $a_t$ which allows solving the
cavity equation \eqref{dota} analytically.  Together with the input-output
relation, the solution provides the transmission and the reflection of the
cavity.

\subsection{Linear response theory}

To obtain the expectation value of the coupling operator, $Z_t$, we assume
that in the absence of the cavity, the system is described by a
density matrix $\rho_0(t)$.  It may refer to any state at
equilibrium or far from equilibrium with a dynamics determined by a
Liouville-von Neumann equation $\dot\rho_0 = \mathcal{L}(t)\rho_0$. The
Liouvillian $\mathcal{L}(t)$ may range from being negligible to cases with
strong time-dependent external forces.  Owing to the interaction with the
cavity, the system experiences an additional driving.  From Eq.~\eqref{Hsc}
with $a$ and $a^\dagger$ replaced by classical amplitudes follows the
corresponding Hamiltonian $H_1(t) = Zf(t)$ with $f(t) = g(a_t+a^*_t)$. Then in
the presence of the cavity, the full master equation of the system becomes
\begin{equation}
\dot\rho = \mathcal{L}(t)\rho -i[Z(t),\rho]f(t) .
\label{ME}
\end{equation}
In an interaction picture that captures all influences but the weak
additional driving, it reads
$\dot{\tilde\rho} = -i[\tilde Z(t),\tilde\rho]f(t)$, where $\tilde
x(t)\equiv \mathcal{U}(t,0)x$ with $\mathcal{U}(t,t')$ the propagator of
the Liouvillian $\mathcal{L}(t)$ which for time-dependent systems may
depend explicitly on both times $t$ and $t'$.

The integrated form of Eq.~\eqref{ME} provides the first-order solution
\begin{equation}
\tilde\rho(t) = \tilde\rho_0(t)
-i\int_{-\infty}^t dt'\, [\tilde Z(t'),\tilde\rho_0(t')] f(t')
\end{equation}
valid for sufficiently weak $H_1(t)$.
Then after a transformation back to the Schr\"odinger picture, the
expectation value of an operator $Z$ at time $t$ becomes
\begin{equation}
\label{lrt}
Z_t
= Z_t^{(0)} + g\int dt'\, \chi(t,t') ( a_{t'} + a^*_{t'})
\end{equation}
with the susceptibility
\begin{equation}
\chi(t,t') = -i\tr \big\{Z \mathcal{U}(t,t')[Z(t'),\rho_0(t')] \big\}
\theta(t-t') \,.
\label{chi}
\end{equation}
Formally, this is the usual Kubo expression, but with the equilibrium
density operator replaced by some non-equilibrium $\rho_0(t')$ which may
depend on the dynamics of the strongly driven qubit as well as on the
initial state.  Henceforth, we focus on the impact of $\rho_0(t)$ on the
cavity.  As the unperturbed expectation value $Z_t^{(0)}$ is
independent of the cavity amplitude $a_t$, it does not contribute to the
frequency shift $\delta\omega$ and, thus, can be neglected.

If the dynamics of the measured system is predominantly coherent, the
propagator of the master equation, $\mathcal{U}$, can be expressed in terms
of the propagator of the Schr\"odinger equation, $U(t,t')$, such that
\begin{equation}
\chi(t,t') = -i\langle[\tilde Z(t,t'),Z]\rangle_0 \theta(t-t') \,,
\label{chicoh}
\end{equation}
with $\tilde Z(t,t') = U^\dagger(t,t') Z U(t,t')$.
The expectation value refers to the unperturbed system density operator
$\rho_0(t')$.  Slow decay of coherent oscillations may still be considered
by a phenomenological decay rate.  This simplified form is already
sufficient for reproducing and generalizing many results from the literature, as
we will see below.  Since the aim of the present work is to highlight the
role of the susceptibility for dispersive readout and not the optimal
computation of this quantity itself, we employ Eq.~\eqref{chicoh} for the
computation of all results, while Eq.~\eqref{chi} will be evaluated in
Appendix~\ref{sec:lindblad} for a particular case to exemplify its use.

\section{Time-independent system}
\label{sec:undriven}

For an undriven $H_\text{sys}$, the susceptibility depends only on the time
difference $t-t'$, such that the $t'$-integration in Eq.~\eqref{lrt} is a
convolution and in frequency space reads $Z_\omega =
g\chi(\omega)(a_\omega + a^*_{-\omega})$.  Consequently, we find the cavity
equation
\begin{equation}
\label{aw}
i(\omega-\omega_0) a_\omega -ig^2\chi(\omega)(a_\omega+a^*_{-\omega}) -\frac{\kappa}{2} a_\omega
 = \sum_{\nu=1,2} \sqrt{\kappa_\nu} a_{\text{in},\nu} .
\end{equation}
For a high-finesse cavity, small detuning $\omega-\omega_0$, and
sufficiently small coupling $g$, such that
\begin{equation}
\label{rwac}
\kappa,\ |\omega-\omega_0|,\ g^2|\chi(\omega)| \ll \omega_0,
\end{equation}
the impact of $a^*_{-\omega}$ is negligible, as is demonstrated in
Appendix~\ref{sec:rwacavity}.  Then the solution of Eq.~\eqref{aw} together
with the input-output relation yields the cavity transmission and
reflection amplitudes at frequency $\omega$,
\begin{align}
\label{tc}
t_c ={}& \frac{a_{\text{out},2}}{a_{\text{in},1}} =
\frac{i\sqrt{\kappa_1\kappa_2}}{\omega_0-\omega+g^2\chi(\omega)-i\kappa/2}
\,,
\\
\label{rc}
r_c ={}& \frac{a_{\text{out},1}}{a_{\text{in},1}}
= 1 + \frac{i\kappa_1}{\omega_0-\omega+g^2\chi(\omega)-i\kappa/2}
\,.
\end{align}
As compared to the absence of the system ($g=0$), the maximum of the
transmission is shifted away from $\omega=\omega_0$ by the ``cavity pull''
\begin{equation}
\delta\omega = g^2 \mathop{\mathrm{Re}}\chi(\omega_0) \,.
\label{deltaw}
\end{equation}
In turn, if the input is monochromatic with $\omega=\omega_0$, the behavior
of $\chi$ becomes manifest in a reduced transmission.  To obtain a
noticeable signal, $\delta\omega$ must be of the order of the cavity line
width.

If $\chi(\omega)$ is real, one readily finds $|t_c|^2+|r_c|^2 = 1$ which
reflects energy conservation. By contrast, the system dissipates energy if
$\mathop{\mathrm{Im}}\chi(\omega) < 0$.  Then, $|t_c|^2+|r_c|^2 < 1$, which
implies energy transfer from the cavity to the system.  Below we will see
that in non-equilibrium situations also the opposite may happen, namely
that the driven system transfers energy to the cavity such that
$|t_c|^2+|r_c|^2>1$.  Nevertheless, we refer to $t_c$ and $r_c$ as
transmission and reflection also in such non-equilibrium situations.

\subsection{Readout of a single qubit}
\label{sec:qubit}

To establish the connection with previous results \cite{Blais2004a,
Zueco2009b}, we turn back to the classic readout of a single qubit with the
Hamiltonian $H_\text{sys} = \frac{\Delta}{2}\sigma_z$ and $Z=\sigma_x$
discussed above.  It is straightforward to obtain the Heisenberg operator
\begin{equation}
\tilde Z(t) = \tilde \sigma_x(t) = \sigma_x\cos(\Delta t) - \sigma_y\sin(\Delta t) ,
\end{equation}
for which Eq.~\eqref{chicoh} is evaluated to read
\begin{equation}
\chi(t)
= 2\langle\sigma_z\rangle \sin(\Delta t) e^{-\gamma t/2} \theta(t)
\label{chiTLS}
\end{equation}
with the phenomenological qubit dephasing rate $\gamma$.  A more profound
calculation may start from Eq.~\eqref{chi} with the dissipative propagator
$\mathcal{U}$ obtained from Bloch-Redfield theory \cite{Redfield1957a,
Blum1996a} or from a Lindblad master equation \cite{Gardiner2004a}.  In
Appendix~\ref{sec:lindblad}, it is shown that for the present example, the
latter also leads to the result in Eq.~\eqref{chiTLS}.

By Fourier transformation $\chi(t)$ turns into
\begin{equation}
\chi(\omega) = \Big(\frac{1}{\Delta-\omega-i\gamma/2}
+\frac{1}{\Delta+\omega+i\gamma/2}\Big)  \langle\sigma_z\rangle ,
\label{qubitshift}
\end{equation}
where for the qubit states $|{\uparrow}\rangle$ and $|{\downarrow}\rangle$
one has $\langle\sigma_z\rangle=\pm1$.  Then the limit $\gamma\to0$ of
$\delta\omega = g^2 \mathop{\mathrm{Re}}\chi(\omega_0)$ is easily
recognized as the non-RWA generalization \cite{Zueco2009b} of the
dispersive shift discussed in Sec.~\ref{sec:model}.  This verifies that for a
single qubit with a time-independent Hamiltonian, dispersive readout
measures the population of the eigenstates.
Interestingly, the presence of qubit dephasing avoids divergences of
$\chi(\omega)$.  Therefore, the second inequality in
Eq.~\eqref{readoutcondition} required for the perturbation theory in
Refs.~\cite{Blais2004a, Zueco2009b} is no longer essential, as long as the
system remains in the linear-response regime.  This can be achieved not
only by a small coupling $g$, but also by reducing the cavity input and,
hence, the additional driving $f(t)$.

A question of practical relevance is the impact of an additional term
$\propto\sigma_x$ such that the system Hamiltonian becomes $H_\text{sys} =
\frac{\Delta}{2}\sigma_z +\frac{\epsilon}{2}\sigma_x$, where in a localized
basis, the additional term corresponds to a detuning of the sites.  The
present formalism provides the answer without performing a technically
involved transformation to the dispersive frame.
The computation of the Heisenberg operator $\tilde Z(t)$ and its commutator with
$Z$ is a straightforward exercise in spin algebra.  After some lines of
calculation, one arrives for $t\geq 0$ at
\begin{equation}
\chi_\epsilon(t) = \frac{2\Delta}{E}\langle\sigma_z\rangle \sin(Et)
+ \frac{2\epsilon\Delta}{E^2}\langle\sigma_y\rangle [1-\cos(Et)] ,
\end{equation}
with the level splitting $E = \sqrt{\Delta^2+\epsilon^2}$.  The first term
is the known expression \eqref{chiTLS}, but now oscillating with angular
frequency $E$ and dressed by a prefactor $\Delta/E$.  The correction given
by the second term vanishes if the system resides in an eigenstate of
$H_\text{sys}$ or $\sigma_z$.  Therefore, we can conclude that the detuning
$\epsilon$ may reduce the sensitivity, but is not a true obstacle for
the readout.

\subsection{Multi-level systems}

Recently, the theory of dispersive readout has been generalized to
multi-level systems to capture the valley degree of freedom in silicon
quantum dots \cite{Burkard2016a, Mi2018a} and the impact of the electron
spin \cite{Petersson2012a, Benito2017a, Samkharadze2018a}.  These works
start from the coupled quantum Langevin equations of the cavity and the
system, which are solved within RWA to obtain the cavity response.

Within the present approach, we employ the weak-dissipation limit of the
susceptibility, Eq.~\eqref{chicoh}, and assume that the initial density
operator is diagonal in the eigenbasis of the system Hamiltonian, i.e.,
$\rho = \sum_{n} p_n|n\rangle\langle n|$, where $H_\text{sys}|n\rangle =
E_n|n\rangle$ with the eigenenergies $E_n$ in ascending order and the populations
$p_n$.  After some lines of algebra we arrive at the expression
\begin{equation}
\chi(\omega) = \sum_{m,n}
\frac{(p_m-p_n)|Z_{mn}|^2}{\omega +E_m-E_n +i\gamma_{mn}/2} \,,
\label{thermal}
\end{equation}
where the level broadenings $\gamma_{mn}$ again have been introduced
phenomenologically.  The generalization to non-diagonal density operators
is straightforward, but beyond the scope of the present work.  A most
relevant special case is a system at thermal equilibrium for which
$\rho\propto\exp(-H_\text{sys}/k_BT)$ is indeed diagonal in the $|n\rangle$
and the probabilities $p_n$ are normalized Boltzmann factors.

Obviously, $\mathop{\mathrm{Re}}\chi(\omega)$ has peaks at $\omega =
E_m-E_n$.  For resonant cavity input ($\omega=\omega_0$), these peaks turn
into dips in the transmission.  As $\chi$ has to be evaluated at $\omega =
\omega_0 >0$, terms with $E_m< E_n$ are off-resonant and smaller than the
ones with interchanged indices.  Consequently, one may neglect the
latter and restrict the summation to terms with $m>n$ to obtain for the
cavity response the RWA result of Ref.~\cite{Benito2017a} [notice that the
$\chi_{mn}$ defined in Refs.~\cite{Burkard2016a, Benito2017a} relate to the
present $\chi(\omega)$ via $g\chi(\omega_0) = \sum_{mn} Z_{mn} \chi_{mn}$].
Equation~\eqref{thermal} generalizes this result beyond RWA.  While the
generalization appears quite intuitive, it has to be stressed that a virtue
of the present approach is the technically easy and transparent way
towards the non-RWA corrections.

A natural demand for dissipative time evolution of a quantum system is that
it preserves the hermiticity of the density operator.  Therefore, the
dephasing rates of density matrix elements must be symmetric in their
index.  The same symmetry holds for the absolute values of the transition
matrix elements $Z_{mn}$.  This has an interesting consequence for the
imaginary part of the expression for $\chi(\omega)$ in Eq.~\eqref{thermal}.
If the populations $p_n$ are a monotonically decaying function of the
energies $E_n$, as is the case at thermal equilibrium, one can readily show
that for $\omega>0$, $\mathop{\mathrm{Im}}\chi(\omega)<0$ (unless all
$\gamma_{mn}=0$ such that $\chi$ becomes real).  Then the system absorbs
energy from the cavity and dissipates it.  Consequently,
$|t_c|^2+|r_c|^2<1$.  In turn, if one establishes by some pumping mechanism
a population inversion, $p_m<p_n$ for at least one pair of states with
$E_m<E_n$, one may find parameter regions with
$\mathop{\mathrm{Im}}\chi(\omega)>0$.  Then the cavity absorbs energy from
the system such that the total cavity output exceeds the cavity input.

\subsection{Relevance of the non-RWA contributions}
\label{sec:rwatest}

To demonstrate the relevance of the non-RWA terms, we first discuss their
impact on the dispersive shift \eqref{qubitshift} for the traditional qubit
readout.  For very weak decoherence, the ratio between the full result and
the RWA result for the cavity pull is readily found as
$\delta\omega/\delta\omega_\mathrm{RWA} \approx
2\epsilon/(\epsilon+\omega_0)$.  Very close to resonance,
$\epsilon\approx\omega_0$, the ratio is close to unity as expected.  In the
vicinity of the resonance, the discrepancy is larger on the flank
with $\epsilon<\omega_0$.

\begin{figure}[tb]
\centerline{\includegraphics{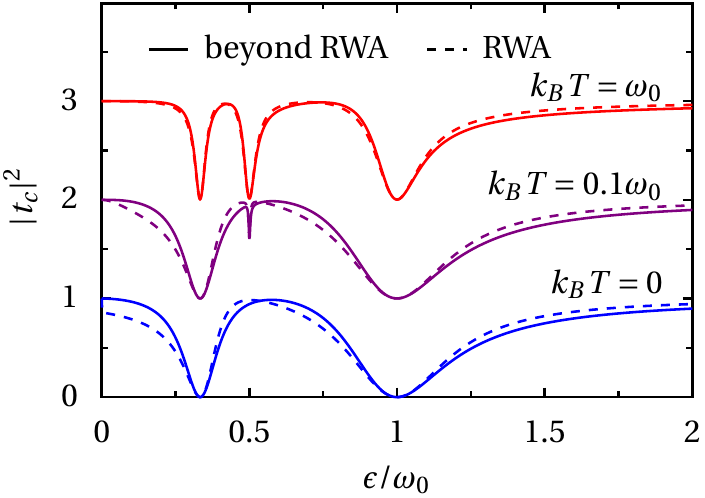}}
\caption{Comparison of the cavity transmission for the system defined in
Eq.~\eqref{H3ls} with the corresponding RWA solution for various
temperatures and all $\gamma_{mn}=0.01\epsilon$.  The cavity is symmetric
$\kappa_1=\kappa_2=\kappa/2$ with $Q=1000$, resonant input
($\omega=\omega_0$), and coupling $g=0.01\omega_0$.  For graphical reasons,
the curves for $k_BT>0$ are shifted vertically.}
\label{fig:testRWA}
\end{figure}

For a closer investigation, we employ a three-level system with the
Hamiltonian and the system-bath coupling given by
\begin{equation}
H_\text{sys} =
\begin{pmatrix} 3\epsilon & 0 & 0 \\ 0 & \epsilon & 0 \\ 0 & 0 & 0 \end{pmatrix} ,
\quad
Z =
\begin{pmatrix} 0 & 1 & 1 \\ 1 & 0 & 1 \\ 1 & 1 & 0 \end{pmatrix} .
\label{H3ls}
\end{equation}
The energies in $H_\text{sys}$ are chosen such that all splittings are
different and, hence, each dip in the cavity transmission can be attributed
easily to a particular transition.

The resulting cavity response for various temperatures is depicted in
Fig.~\ref{fig:testRWA}.  It demonstrates that far from any resonance, the cavity
transmission is perfect, $|t_c|^2=1$, while dips may emerge when an energy
splitting of $H_\text{sys}$ matches the cavity frequency.  At zero
temperature, the transition between the first and the second excited state
remains dark.  Equation~\eqref{thermal} explains this fact by the vanishing
populations $p_2=p_3=0$. Moreover, one notices that beyond RWA, the
asymmetry of the dip increases, as is expected from the introductory discussion
of a single qubit.  While the difference between the RWA and the non-RWA
solution is moderate, the non-RWA terms have a clear impact on the shape of
the dips which may be relevant for quantitative comparisons between
experiment and theory.

With increasing temperature, the excited states become thermally populated
and, thus, a further dip shows up for $2\epsilon=\omega_0$.  Once the
temperature is of the order of the splittings, all states have similar
population and the numerator in Eq.~\eqref{thermal} becomes small.  Then
the cavity response becomes weaker, which is visible in the reduced line
width.  This implies that the dips have less overlap and are no longer
affected by their neighbors.  Since the non-RWA terms formally correspond
to peaks at negative frequencies, the quality of a RWA is expected to
improves when the signal is weaker.  The same holds true when for smaller
system-cavity coupling, the width of the dips shrinks (not shown).

\section{AC-driven system}
\label{sec:ac}

For a periodically time-dependent $H_\text{sys}(t) = H_\text{sys}(t+T)$
with driving frequency $\Omega = 2\pi/T$, the cavity response has been
derived in Ref.~\cite{Kohler2017a}.  Here we present details of the
derivation and discuss the relation to the undriven case and the
difficulties with establishing a RWA.  The application of the formalism to
specific situations emphasizes its usefulness for solid-state quantum
information processing.

For time-dependent systems, the evaluation of Eqs.~\eqref{chi} and
\eqref{chicoh} is hindered by the fact that the susceptibility $\chi(t,t')$
generally depends explicitly on both times.  Nevertheless time-periodicity
allows a simplification in the long-time limit, because after a transient
stage, the $T$-periodicity of the Hamiltonian leads to $\chi(t,t') =
\chi(t+T,t'+T)$ \cite{Kohler2005a}.  Therefore, introducing the time
difference $\tau = t-t'$ allows one to conclude that $\chi(t,t-\tau)$ is
$T$-periodic in $t$, such that it can be written as a combination of
Fourier series and integral,
\begin{equation}
\label{chikw}
\chi(t,t-\tau) = \sum_k\int \frac{d\omega}{2\pi}\,
e^{-ik\Omega t-i\omega\tau} \chi^{(k)}(\omega) \,.
\end{equation}
Then the Fourier representation of the system response $Z_t$ in
Eq.~\eqref{lrt} becomes
\begin{equation}
Z_\omega = g \sum_k \chi^{(k)}(\omega-k\Omega)(a_{\omega-k\Omega}
+a^*_{-\omega+k\Omega}).
\end{equation}
The summation over the sideband index $k$ reflects the frequency mixing
inherent in the linear response of a driven quantum system.

To proceed, we restrict ourselves to the limit in which dispersive readout is
usually performed, i.e., to a resonantly driven high-finesse cavity with
$\kappa\ll\omega_0\approx\omega$.  Then for $\Omega\gtrsim\kappa$, all
$a_{\omega-k\Omega}$ with $k\neq 0$ will be outside the cavity linewidth
and, hence, can be neglected.
As in the undriven case, the complex conjugate mode $a_{-\omega+k\Omega}^*$
for $k=0$ and $\omega\approx\omega_0$ will be far off resonance and, thus,
will not be excited.  Nevertheless, one sideband may be in resonance with
$a_{\omega_0}$ if the difference of the frequencies $\omega_0$ and any
$-\omega_0+k\Omega$ is smaller than the cavity linewidth.  For a
high-finesse cavity, such frequency matching is unlikely and
already a tiny deviation from the resonance $k\Omega = 2\omega_0$ leads to
a time-dependent phase factor in $a_{-\omega_0+k\Omega}^*$.  In an
experiment, the phase may even drift and be practically random.  Therefore
we assume that we can continue with a phase-average in which
$a_{-\omega_0+k\Omega}^*$ vanishes.  Then the system response becomes
$Z_\omega = g\chi^{(0)}(\omega) a_\omega$.  Continuing as in
Sec.~\ref{sec:undriven}, we again obtain the transmission and reflection
amplitudes \eqref{tc} and \eqref{rc}, but with the replacement
\begin{equation}
\chi(\omega) \longrightarrow \chi^{(0)}(\omega) .
\label{chi-chi0}
\end{equation}
Thus, we have demonstrated that in the decomposition \eqref{chikw} of the
susceptibility, the component relevant for dispersive readout is the one
with $k=0$, i.e., the one that corresponds to the $t$-average of
$\chi(t,t-\tau)$.

The remaining computation of $\chi^{(0)}(\omega)$ may be performed
with the Floquet-Bloch-Redfield formalism developed in
Ref.~\cite{Kohler1997a}.  It starts by diagonalizing
$H_\text{sys}(t)-i\partial_t$ in the Hilbert space extended by the space of
$T$-periodic functions \cite{Shirley1965a, Sambe1973a} to obtain the
Floquet states $|\phi_\alpha(t)\rangle = |\phi_\alpha(t+T)\rangle$, the
quasienergies $\epsilon_\alpha$ 
and the stationary solutions of the Schr\"odinger equation,
$|\psi_\alpha(t)\rangle = e^{-i\epsilon_\alpha t}|\phi_\alpha(t)\rangle$.
The corresponding expression for the propagator, $U(t,t') = \sum_\alpha
e^{-i\epsilon_\alpha(t-t')} |\phi_\alpha(t)\rangle
\langle\phi_\alpha(t')|$, allows us to deal with the interaction picture
operators in $\chi(t,t')$.

As in the undriven case, we restrict ourselves to the limit of weak
decoherence and assume that the susceptibility can be written in the form
of Eq.~\eqref{chicoh}.  Moreover, it is known \cite{Kohler1997a} that for
very weak dissipation, the long-time solution of an ac-driven quantum
system becomes diagonal in the Floquet basis.  Hence, $\rho_\text{sys}(t) =
\sum p_\alpha|\phi_\alpha(t)\rangle \langle\phi_\alpha(t)|$ with $p_\alpha$
the occupation probabilities of the Floquet states computed as described in
Appendix~\ref{sec:FBR}.  Notice that frequently, one refers to the diagonal
approximation of the density operator also as RWA, which however must be
distinguished from the RWA discussed here.  With these ingredients, we find
\begin{equation}
\label{chi0w}
\chi^{(0)}(\omega) = \sum_{\alpha,\beta,k}
\frac{(p_\alpha-p_\beta)|Z_{\alpha\beta,k}|^2}
{\omega+\epsilon_\alpha-\epsilon_\beta-k\Omega+i\gamma_{\alpha\beta}/2},
\end{equation}
where $Z_{\alpha\beta,k}$ denotes the $k$th Fourier component of the
$T$-periodic transition matrix element $Z_{\alpha\beta}(t) =
\langle\phi_\alpha(t)|Z|\phi_\beta(t)\rangle$.  Once more, the dephasing
rate $\gamma_{\alpha\beta}$ has been introduced phenomenologically.

An important observation is now that one expects a signal in the
cavity transmission when the denominator in Eq.~\eqref{chi0w} assumes its
minimum, i.e., when the real part of $\chi^{(0)}(\omega)$ vanishes.  For a
resonantly driven cavity this is the case for
\begin{equation}
\epsilon_\alpha-\epsilon_\beta = \omega_0 +k\Omega \,.
\label{ACresonance}
\end{equation}
While Eq.~\eqref{thermal} predicts for time-independent systems a signal
when the oscillator frequency matches an energy difference, we obtain the
natural generalization to ac-driven systems, namely that energies are
replaced by quasienergies shifted by multiples of the driving frequency
$\Omega$.

The presence of $k\Omega$ in Eq.~\eqref{chi0w} represents a difficulty
for establishing a RWA, because the terms with $k<0$ invalidate the
arguments employed in the undriven case.  A further obstacle is the Brillouin
zone structure of the quasienergies \cite{Shirley1965a, Sambe1973a} which
even does not allow a proper ordering or a direct relation between the
quasienergies and the populations.  Therefore, one generally is forced to work beyond
RWA.  In the limit of adiabatically slow driving, one nevertheless finds
an expression that resembles a RWA solution \cite{Kohler2017a,
Mi2018b}.  Its physical origin, however, is different.


\subsection{Cavity-assisted LZSM interference}
\label{sec:lzsm}

\begin{figure}
\centerline{\includegraphics{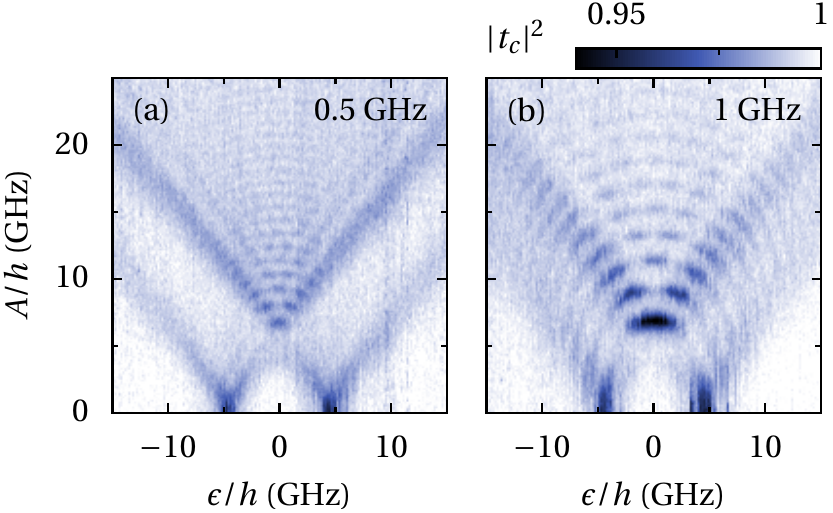}}
\caption{Experimental transmission of a microwave resonator coupled to an
ac-driven GaAs double quantum dot reported in Figs.~2(c,d) of
Ref.~\cite{Koski2018a}.  The driving frequencies are $\Omega/2\pi=0.5\,$GHz
(a) and $\Omega/2\pi=1\,$GHz (b), while tunnel matrix element and decoherence
rate read $\Delta/h=6.8\,$GHz and $\gamma/2\pi=400\,$MHz, respectively.  The
cavity with bare frequency $\omega_0/2\pi = 8.32\,$GHz and decay rate
$\kappa/2\pi=110\,$MHz couples to the qubit with strength $g/h=30\,$MHz.
}
\label{fig:lzsmexp}
\end{figure}

\begin{figure*}
\centerline{\includegraphics{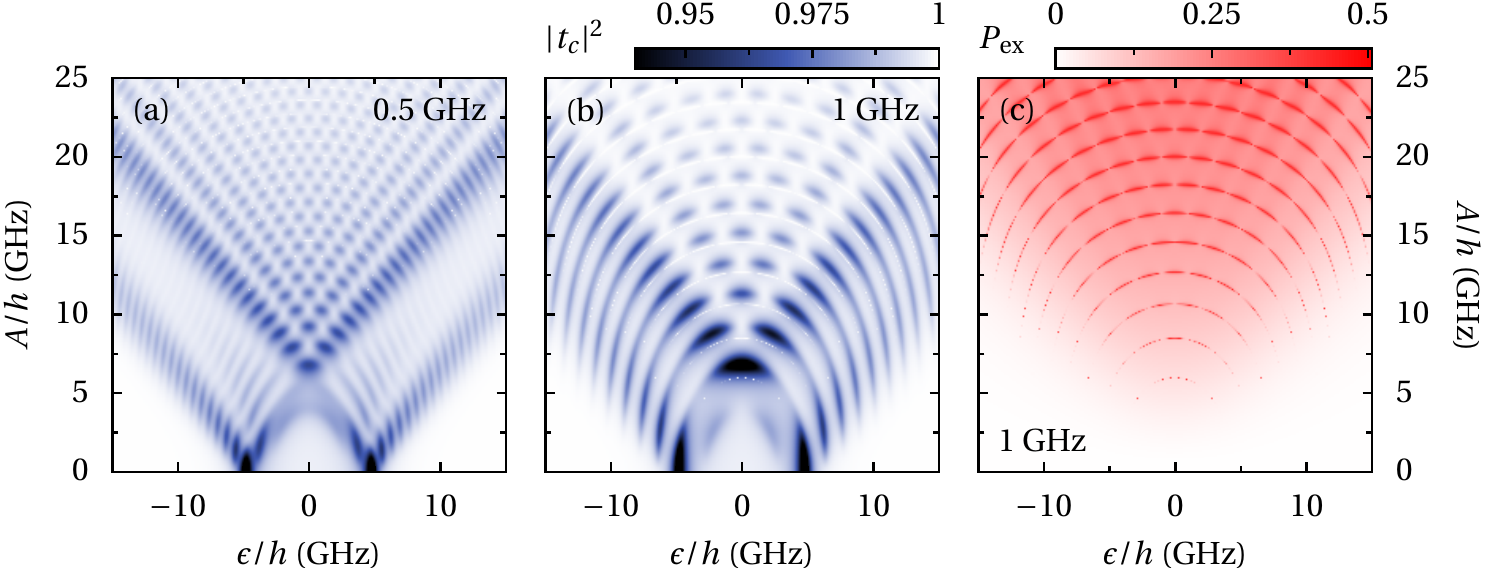}}
\caption{(a,b) Theory data for the cavity-assisted LZSM interference
pattern shown in Fig.~\ref{fig:lzsmexp}
computed with the susceptibility in Eq.~\eqref{chi0w}.
(c) Mean population of the excited state of $H_0$ for the parameters used
in panel (b).}
\label{fig:lzsm}
\end{figure*}

As a first example, we investigate Landau-Zener-St\"uckelberg-Majorana
(LZSM) interference \cite{Shevchenko2010a} which occurs when a qubit is
repeatedly swept through an avoided crossing that acts like a beam
splitter.  The resulting interference patterns as a function of the average
detuning and the driving amplitude have been used to demonstrate the
coherence of qubits \cite{Oliver2005a, Sillanpaa2006a, DupontFerrier2013a,
Stehlik2012a} and to determine the coupling of a charge qubit to a
dissipating environment \cite{Forster2014a}.  To be specific, we consider
the time-dependent Hamiltonian
\begin{equation}
H_\text{sys}(t) = \frac{\Delta}{2}\sigma_x + \frac{\epsilon+A\cos(\Omega
t)}{2} \sigma_z .
\end{equation}
In contrast to Sec.~\ref{sec:qubit}, the pseudo-spin operators $\sigma_i$
here are represented in the basis of localized states such that the dipole
coupling between system and cavity is established by the operator $Z =
\sigma_z$.  The Hamiltonian in the absence of the driving ($A=0$) will be
denoted by $H_0$.  The populations of the Floquet states are computed with
a system-bath coupling via $\sigma_x$, see Appendix~\ref{sec:FBR}.

Recently, this system including the cavity has been employed for an
experimental demonstration of low-frequency LZSM patterns in the cavity
transmission \cite{Koski2018a}.  Figure \ref{fig:lzsmexp} depicts two
measured pattern, while the corresponding theoretical results obtained with
Eqs.~\eqref{tc}, \eqref{chi-chi0}, and \eqref{chi0w} are plotted in
Fig.~\ref{fig:lzsm}(a,b).  Theory and experiment exhibit a striking
quantitative agreement.  Moreover, the resonance condition for the location
of the fringes conjectured in Ref.~\cite{Koski2018a} agrees with
Eq.~\eqref{ACresonance}, which means that it can be derived from the
present theory for dispersive readout of ac-driven quantum systems.

Figure \ref{fig:lzsm}(c) depicts the time-averaged non-equilibrium
population of the excited state of $H_0$.  This quantity also exhibits a
LZSM pattern which, however, is remarkably different from the one for the
transmission.  First, the pronounced structure close to the bisecting lines
$A\approx \pm\epsilon$ is absent.  Second, the
interference fringes appear at different positions, which becomes
particularly evident when one pays attention to their nodes.  The resonance
conditions provide an explanation for the discrepancy.  A fringe in the
transmission requires Eq.~\eqref{ACresonance} be fulfilled, while the
corresponding expression for the non-equilibrium population does not
contain the cavity frequency $\omega_0$ \cite{Shevchenko2010a}.  This
implies that for an ac-driven qubit---in contrast to the undriven one in
Sec.~\ref{sec:qubit}---the signal of dispersive readout does not
necessarily reflect the population of the excited state.  Nevertheless,
patterns of the readout signal such as those shown in
Figs.~\ref{fig:lzsm}(a,b) can be explained in terms of repeated
Landau-Zener transitions, but between qubit states dressed by the cavity
mode \cite{Mi2018b}.  This idea of cavity-assisted LZSM interference
qualitatively reproduces the structure of measured patterns if one replaces
in the low-frequency theory of Ref.~\cite{Shevchenko2010a} the qubit states
by dressed states, as has been demonstrated with qubits \cite{Koski2018a}
as well as with multi-level systems \cite{Mi2018b}.

\subsection{Two-tone spectroscopy}

\begin{figure}[tb]
\centerline{\includegraphics{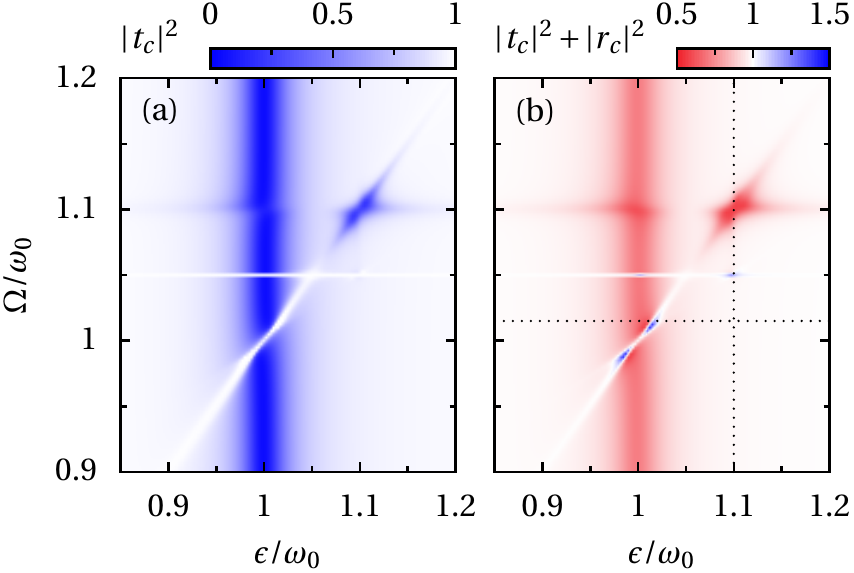}}
\caption{Two-tone spectroscopy of the three level system defined in
Eq.~\eqref{Htt} with $E_0=0$, $E_2=2.1\,\omega_0$, driving amplitude
$A=0.01\omega_0$, and system-cavity coupling $g=0.003\,\omega_0$.
(a) Cavity transmission $|t_c|^2$.
(b) Sum of transmission and reflection, $|t_c|^2+|r_c|^2$, demonstrating
the energy absorption (red) and emission (blue) by the driven qubit.
All other parameters are as in Fig.~\ref{fig:testRWA}.
Dotted lines mark values used in Fig.~\ref{fig:twotone}.
\label{fig:twotone2D}
}
\end{figure}

\begin{figure}[tb]
\centerline{\includegraphics{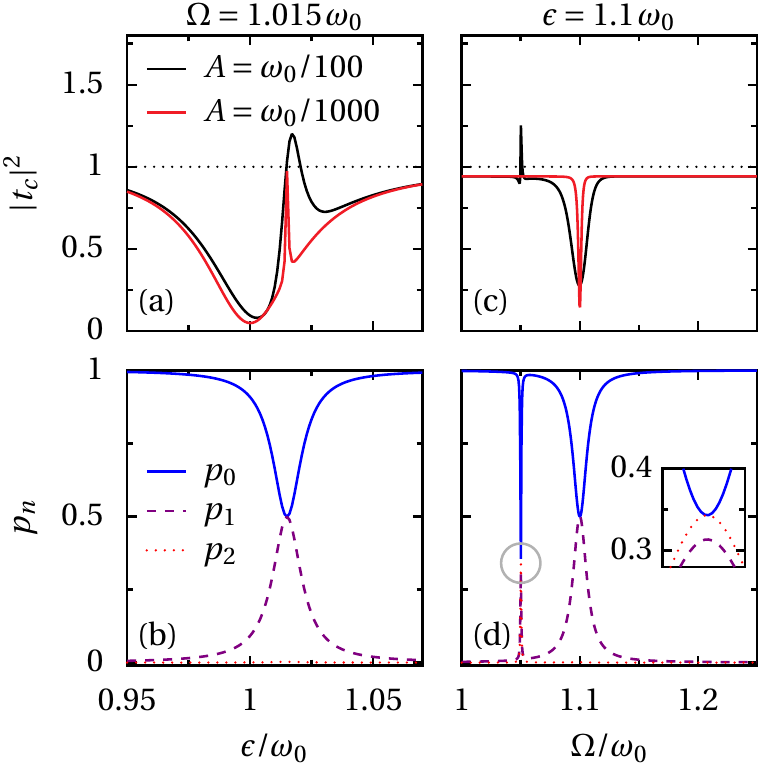}}
\caption{(a) Cavity transmission for two-tone spectroscopy at constant driving
frequency as a function of the energy of state $|1\rangle$.
All other parameters are as in Fig.~\ref{fig:twotone2D}.
(b) Average populations of the eigenstates of $H_0$ for
driving amplitude $A=0.01\omega_0$ corresponding to the black line in
panel (a).
(c,d) The same for constant energy of state $|1\rangle$ as a function of
the driving frequency.  The inset is an enlargement of the region marked in
the main panel by a grey circle.  It demonstrates the population inversion
between states $|2\rangle$ (red dotted) and $|1\rangle$ (purple, dashed) at
the two-photon resonance $E_2-E_0 = 2\Omega$.  For significantly smaller
amplitudes, the inversion vanishes.
\label{fig:twotone}
}
\end{figure}

When the driving frequency $\Omega$ is of the order of the cavity frequency
$\omega_0$, interesting effects emerge already for relatively small
amplitudes.  For example, the driving may induce transitions from the
ground state to excited states and, thus, affect the populations.  The
consequences of such ground state depletion can be understood qualitatively
already from the susceptibility for the undriven situation,
Eq.~\eqref{thermal}, while Floquet theory serves for a quantitative
prediction of effects of higher order in the amplitude.  These effects have
similarities with pump-probe spectroscopy \cite{Fischer2016a} despite that
the second driving is not pulsed.

Let us therefore investigate the three-level system of
Sec.~\ref{sec:rwatest} with an ac driving and with the energies of the
ground state and the second excited state kept at the constant values.
Then the system Hamiltonian reads
\begin{equation}
H_\text{sys}(t) =
\begin{pmatrix} E_2 & 0 & 0 \\ 0 & \epsilon & 0 \\ 0 & 0 & E_0 \end{pmatrix}
+ A Z \cos(\Omega t)
\label{Htt}
\end{equation}
with the operator $Z$ and the system-bath coupling as above.  For
simplicity, we restrict the discussion to rather low temperatures at which
thermal excitations do not play a role.  The amplitude is chosen moderately
large such that effects of higher order in $A$ start to play a role, but do
not dominate.

Figure~\ref{fig:twotone2D}(a) depicts the cavity transmission as a function
of the energy splitting between the two lowest states, $\epsilon-E_0$.
Its structure is governed by the various resonances of the
system.  The dominating one is visible as a broad vertical line when the
lowest system transition matches the cavity frequency, $\epsilon =
\omega_0$.
Furthermore, the ac driving may deplete the ground state by inducing transitions to
the states $|1\rangle$ and $|2\rangle$.  For relatively small amplitudes, this
happens when a condition $E_{1,2}-E_0 = k\Omega$, $k=1,2,\ldots$,
is met, where $k>1$ corresponds to multi-photon resonances which have
smaller impact.  The resulting depletion of state $|0\rangle$ can be
appreciated as white lines at $E_1-E_0=\Omega$ and at $E_2-E_0=2\Omega$.
The corresponding populations of the eigenstates of $H_0$ shown in
Figs.~\ref{fig:twotone}(b,d) confirm the natural expectation that only the
resonance conditions involving $E_2$ induce the excitations to state
$|2\rangle$.  A particular feature is the resonance island when the conditions
$E_1-E_0=\Omega$ and $E_2-E_1 = \omega_0$ are simultaneously fulfilled.  Then the
driving creates a significant population of state $|1\rangle$, while the
cavity probes the transition from $|1\rangle$ to $|2\rangle$.  For small
amplitudes, this leads to very sharp lines that may be used for calibration
\cite{Samkharadze2018a, Mi2018b}, as can be appreciated from the red line in
Fig.~\ref{fig:twotone}(c).

In most parameter regions in which the cavity response is sensitive to the
system, the system absorbs energy from the cavity.  Then the sum of
transmission and reflection is smaller than unity, see
Fig.~\ref{fig:twotone2D}(b).  There exist, however, also small regions in
which the driven quantum systems transfers energy to the cavity such that
$|t_c|^2+|r_c|^2>1$.  This effect may be explained by population inversion
stemming from an interplay of driving and dissipation \cite{Ferron2012a,
Blattmann2015a}; see inset of Fig.~\ref{fig:twotone}(d).  However,
Figs.~\ref{fig:twotone}(a) and \ref{fig:twotone}(b) demonstrate that
at $\Omega \approx\epsilon \approx 1.015\,\omega_0$ an energy transfer to the
cavity is possible even in the absence of a population inversion.
The reason for this is that sidebands in the susceptibility
\eqref{chi0w} may give rise to $\mathop{\mathrm{Im}}\chi(\omega_0)>0$
irrespective of the sign of $p_\alpha-p_\beta$.  For smaller driving
amplitudes, the impact of the driving is reduced and eventually the
imaginary part of $\chi$ becomes again negative such that $|t_c|^2+|r_c|^2$
is bounded by unity as in the undriven case.


\section{Discussion and conclusions}
\label{sec:summary}

We have developed a versatile theory for dispersive readout based on a
relation between the cavity response and a susceptibility of the system to
be measured.  It holds in and out of equilibrium and reveals that
dispersive readout detects the autocorrelation of the system operator by
which the coupling to the cavity is established.  Besides being of
appealing generality, the approach enables straightforward calculations
with moderate effort, in particular the generalization of previous results
beyond a rotating-wave approximation to the system and for the treatment of
time-dependent systems.

To demonstrate these virtues, we have reproduced in a technically
effortless way the result for qubit readout beyond a rotating-wave
approximation for the measured system \cite{Zueco2009b}
without the need for a rather involved transformation to the dispersive
frame.  Moreover, we have generalized it to qubit Hamiltonians
that include detuning.  For the readout of multi-level systems, the
non-rotating-wave
corrections turned out to play a role at low temperatures and for strong
system-cavity coupling.  These corrections essentially lead to asymmetries
in the transmission peaks which may be relevant for the agreement with
experimental results.

For the readout of ac-driven systems, we have provided details of the
Floquet approach of Ref.~\cite{Kohler2017a} and have found that the
relevant component of the response function can be interpreted as
time-averaged susceptibility.  As the sidebands of Floquet states
correspond to components with different energies, the common line of
argumentation towards a rotating-wave approximation becomes invalid.
The application to strongly driven qubits that undergo cavity-assisted
Landau-Zener-St\"uckelberg-Majorana
interference shows a striking agreement with recent experimental results
\cite{Koski2018a} which emphasizes the suitability of the formalism.
As a further test case, we have considered two-tone spectroscopy which can
be employed for the calibration of level splittings.  The present approach
not only confirmed features that can be deduced qualitatively from the
formula for the undriven case.  In particular for intermediate amplitudes,
it also predicts less evident features such as the energy transfer from the
driven system to the cavity.

The cases studied consider rather weak dephasing that can be
described by exponentially decaying phase factors.  For stronger
dissipation the treatment of the cavity still holds, while the
susceptibility may have to be computed with more elaborated techniques
\cite{Weiss1999a, Breuer2003a}.  Recently, such techniques have been
employed for describing the direct probe of a superconducting quantum circuit
\cite{Magazzu2018a}.

\begin{acknowledgments}
I would like to thank Andr\'as P\'alyi, Jonne Koski, and M\'onica Benito
for discussions and for carefully reading the manuscript.  Moreover, I am
grateful to the authors of Ref.~\cite{Koski2018a} for providing me with the
experimental data depicted in Fig.~\ref{fig:lzsmexp}.
This work was supported by the Spanish Ministry of Economy and
Competitiveness via Grant No.\ MAT2017-86717-P.
\end{acknowledgments}

\appendix

\section{Qubit susceptibility from Lindblad theory}
\label{sec:lindblad}

As an example for the direct evaluation of $\chi$ from Eq.~\eqref{chi},
we consider the qubit readout discussed in Sec.~\ref{sec:qubit}
for the initial state $\rho(t') = \frac{1}{2}\mathbf{1} + \vec p\cdot
\vec\sigma$ with the Bloch vector $\vec p =\frac{1}{2}
\langle\vec\sigma\rangle$.  Then for $Z=\sigma_x$, the commutator in
Eq.~\eqref{chi} becomes $2i p_y\sigma_z - 2ip_z\sigma_y$.

The dissipative qubit dynamics is assumed to be governed by the Markovian
master equation $\dot\rho = \mathcal{L}\rho$ with the Lindblad
superoperator \cite{Gardiner2004a}
\begin{equation}
\mathcal{L}\rho = -\frac{i\epsilon}{2}[\sigma_z,\rho]
+\frac{\gamma}{2}(2\sigma_-\rho\sigma_+ -\sigma_+\sigma_-\rho
-\rho\sigma_+\sigma_-) \,,
\end{equation}
where $\sigma_i$ denote the usual Pauli matrices.  It is straightforward to
show that the master equation possesses the eigensolutions
\begin{equation}
\sigma_-\sigma_+ , \quad
\sigma_z e^{-\gamma t},\quad
\sigma_+ e^{-i\epsilon t-\gamma t/2},\quad
\sigma_- e^{i\epsilon t-\gamma t/2} ,
\label{app:eigensolutions}
\end{equation}
the first one being the equilibrium solution
$|{\downarrow}\rangle\langle{\downarrow}|$.  Together with the
eigensolutions of the adjoint superoperator, one may construct the
propagator and evaluate the expression for $\chi$.  For the present case
an elegant shortcut exists, because the first two eigensolutions are
diagonal and vanish after multiplication with $Z=\sigma_x$ and taking the
trace.  Therefore only the term $-2ip_z \sigma_y =
\langle\sigma_z\rangle(\sigma_--\sigma_+)$ will contribute.  Upon inserting
the time evolution of $\sigma_\pm$ given by the third and the fourth
eigensolution, one readily finds expression \eqref{chiTLS}.  Beyond
Lindblad, e.g., for a Bloch-Redfield master equation, the calculation
becomes more involved, but conceptually follows the same lines.

\section{RWA to the cavity mode}
\label{sec:rwacavity}

While a central issue of this work is the treatment of the system response
function $\chi$ beyond RWA, neglecting in Eq.~\eqref{aw} the contribution
with $a^*_{-\omega}$ represents a RWA for the cavity mode.  In the
following, we derive the conditions under which this approximation holds.

Equation \eqref{aw} for the cavity amplitude $a_\omega$ together with the
corresponding equation for $a^*_{-\omega}$ forms a closed set of
linear equations,
\begin{equation}
M \begin{pmatrix} a_\omega \\ a^*_{-\omega} \end{pmatrix}
= i\sum_{\nu=1,2} \sqrt{\kappa_\nu}
  \begin{pmatrix} a_{\text{in},\nu}(\omega) \\ a^*_{\text{in},\nu}(-\omega) \end{pmatrix}
\label{Ma}
\end{equation}
with the matrix
\begin{equation}
M = \begin{pmatrix} A(\omega) & g^2\chi(\omega) \\
g^2\chi^*(-\omega) & A^*(-\omega)
\end{pmatrix}
\end{equation}
and $A(\omega) = \omega_0-\omega +g^2\chi(\omega) -i\kappa/2$.
In principle, Eq.~\eqref{Ma} can be solved for $a_\omega$ exactly, but the
resulting expressions are not very concise.  A simplification can be
achieved under the conditions in Eq.~\eqref{rwac} which physically
correspond to the following situation.  To obtain a reasonably strong
signal, the cavity, must have a large $Q = \omega_0/\kappa \gg 1$ and must
be driven close to resonance such that $|\omega-\omega_0| \ll \omega_0$.
Moreover, even in the strong-coupling limit, the dispersive shift
$g^2\mathop{\mathrm{Re}}\chi(\omega)$ is much smaller than the bare cavity
frequency.  Then the inverse of the matrix $M$ is approximately given by
\begin{equation}
M^{-1} \approx \begin{pmatrix} A(\omega)^{-1} & 0 \\ 0 & 0
\end{pmatrix} ,
\end{equation}
where the corrections are of higher order in the small frequencies
on the left-hand side of Eq.~\eqref{rwac}.  Computing $a_\omega$ with this
expression for $M^{-1}$ is equivalent to ignoring $a^*_{-\omega}$ in
Eq.~\eqref{aw}.

The same result can be obtained by assuming a monochromatic cavity input
$a_{\text{in},1}(\omega) \propto \delta(\omega-\omega_r)$ with $\omega_r
\approx \omega_0$.  Then $a^*_{\text{in},1}(-\omega)$ vanishes and one finds
from Eq.~\eqref{Ma} that $a^*_{-\omega}$ is smaller than
$a_\omega$ roughly by a factor $g^2\chi^*(-\omega)/2\omega_0$.

\section{Floquet-Bloch-Redfield theory}
\label{sec:FBR}

In Sec.~\ref{sec:ac}, the occupation probabilities of the Floquet states
are computed with the approach derived in Ref.~\cite{Kohler1997a}.  Its
starts from a system-bath model in which the driven quantum system is
coupled to an ensemble of harmonic oscillator, $H_\text{bath} = \sum_\nu
\omega_\nu b_\nu^\dagger b_\nu$, and the interaction Hamiltonian
$H_\text{sys-bath} = X\sum_\nu \lambda_\nu(b_\nu^\dagger+b_\nu)$.  The
influence of the bath is determined by the spectral density $J(\omega) =
\pi\sum_\nu |\lambda_\nu|^2 \delta(\omega-\omega_\nu) \equiv
\pi\alpha\omega/2$ with the dimensionless dissipation strength~$\alpha$.

To treat this model, we employ the Bloch-Redfield master equation
\cite{Redfield1957a, Blum1996a} decomposed into the Floquet basis in which
for $\alpha\ll1$, it eventually becomes diagonal \cite{Kohler1997a}.  This
motivates for the long-time solution the ansatz $\rho(t) = \sum_\alpha
|\phi_\alpha(t)\rangle \langle\phi_\alpha(t)|$ and leads to the Pauli-type
master equation
\begin{equation}
\dot p_\alpha = \sum_\beta (w_{\alpha\leftarrow\beta}p_\beta
-w_{\beta\leftarrow\alpha} p_\alpha)
\end{equation}
for the populations $p_\alpha$.  The transition rates are conveniently
expressed in terms of the Fourier components of the Floquet states,
$|\phi_{\alpha,k}\rangle$ defined implicitly by the Fourier series
$|\phi_\alpha(t)\rangle = \sum_k e^{-ik\Omega t}|\phi_{\alpha,k}\rangle$.  After
some algebra one obtains
\begin{equation}
\label{FBR}
w_{\alpha\leftarrow\beta} = 2\sum_k \Big| \sum_{k'} \langle
\phi_{\alpha,k'+k}|X|\phi_{\beta,k'}\rangle \Big|^2
N(\epsilon_\alpha-\epsilon_\beta+k\Omega)
\end{equation}
with $N(\epsilon) = J(\epsilon)n_\text{th}(\epsilon)$ and the bosonic
thermal occupation number $n_\text{th}(\epsilon) =
[\exp(\epsilon/k_BT)-1]^{-1}$.  Here, for negative energies, the spectral density
of the bath is defined as $J(-\epsilon)=-J(\epsilon)$ and $n_\text{th}$
follows by analytic continuation.
Notice that the long-time solution of the master equation \eqref{FBR} is
independent of the dissipation strength $\alpha$, but consistency of the
approach requires $\alpha\ll1$.  The data in Sec.~\ref{sec:ac} are computed
for a system-bath model with $X=\sigma_x$ which yields LZSM patterns with a
generic shape that is robust against small variations of the coupling operator
\cite{Blattmann2015a}.

%

\end{document}